# Exponentially confining potential well

A. D. Alhaidari

*Saudi Center for Theoretical Physics, P.O. Box 32741, Jeddah 21438, Saudi Arabia*

**Abstract:** We introduce an exponentially confining potential well that could be used as a model to describe the structure of a strongly localized system. We obtain an approximate partial solution of the Schrödinger equation with this potential well where we find the lowest energy spectrum and corresponding wavefunctions. The Tridiagonal Representation Approach (TRA) is used as the method of solution, which is obtained as a finite series of square integrable functions written in terms of the Bessel polynomial.

*To the memory of my friend, colleague and collaborator, the late*

***Mohammed S. Abdelmonem.***

## 1. Introduction

Confining potentials are used as models to describe the structure of bound systems with strong localization. The harmonic oscillator potential, which is treated in most textbooks on quantum mechanics, is the most popular (see, for example, Ref. [1]). Other models include the linear potential, which is sometimes used to describe the confinement of quarks inside hadrons [2,3]. These two models describe confinement of a quadratic and linear strength, respectively. Aside from the infinite square well and the quartic potentials, there is rarely any treatment in the literature of potentials with extreme confinement strength including those of the exponential type. If one is interested in models of extreme confinement but without infinitely hard boundaries (to allow for some level of wall penetration and non-vanishing of the wavefunction tail into the walls) then one cannot use the infinite square well potential and must have confinement strength better than power-law. In this study, we provide such a model in one dimension over the whole real line where the strength of confinement grows exponential. Specifically, we propose the following potential model

$$V(x) = \frac{\lambda^2}{2}\left(\tfrac{1}{4}e^{-2\lambda x} - A_- e^{-\lambda x} + A_+ e^{\lambda x}\right), \qquad (1)$$

where $-\infty < x < +\infty$ and we have adopted the atomic units $\hbar = M = 1$. The scale parameter $\lambda$ is real with inverse length dimension, which is a measure of the range of the potential. The parameters $A_\pm$ are real and dimensionless with $A_+ > 0$. Figure 1 is a plot of this potential for a fixed $A_+$ and several values of $A_-$. Changing the sign of $\lambda$ will cause reflection of the plot with respect to the vertical axis. Zooming out along the vertical axis in the figure (i.e., for large energies), the potential mimics an infinite square well.

The Schrödinger equation with the potential function (1) is not exactly solvable using any of the conventional methods of solutions unless $A_+ = 0$ in which case (1) becomes the Morse potential. Such conventional methods include factorizations, point canonical transformation, supersymmetry, shape invariance, Darboux transformation, second quantization, asymptotic iteration method, group theory, path integral transformation, Nikiforov-Uvarov method, etc. (for description of these methods, see as examples, Ref. [4-11]). Nonetheless, we will demonstrate here that the newly introduced algebraic method of solution known as the "Tridiagonal Representation Approach (TRA)" [12] will give a quasi-exact solution of this problem in the form of a bounded series of square integrable functions. The expansion coefficients of the series are orthogonal polynomials in the energy and physical parameter space.

We start by writing the solution of the Schrödinger equation with the potential (1) as the following series

$$\psi(x) = \sum_n f_n \phi_n(x), \qquad (2)$$

where $\{\phi_n(x)\}$ is a complete set of square integrable functions that vanish at infinity and $\{f_n\}$ are the expansion coefficients. Making a coordinate transformation to the dimensionless variable $y(x) = e^{\lambda x}$, the resulting time-independent Schrödinger equation suggests that we choose the basis set with the following elements

$$\phi_n(x) = G_n y^\alpha e^{-\beta/y} J_n^\mu(y), \qquad (3)$$

where $J_n^\mu(y)$ is the Bessel polynomial whose properties are given in Appendix A. The parameter $\mu$ is negative such that $\mu < -N - \frac{1}{2}$, where $N$ is a non-negative integer and $n = 0,1,2,..,N$. The dimensionless real parameters $\alpha$ and $\beta$ will be determined below whereas the normalization constant is conveniently chosen as $G_n = \sqrt{\frac{-(2n+2\mu+1)}{n!\Gamma(-n-2\mu)}}$.

Our choice of basis (3) is suggested by the TRA requirement. We could not come up with an alternative basis set that result in a tridiagonal matrix representation for the wave operator with the potential (1). Unfortunately, this basis set is not complete. Physically, completeness of a *discrete* basis set means that (i) its size is infinite, (ii) it is defined over the whole configuration space of the physical problem, and (iii) it satisfies the boundary conditions. The last requirement usually implies that the basis elements are square integrable. Rigorously, one should also require that the basis set be dense over the whole configuration space. For example, deleting some finite subset of the infinite basis will render it incomplete. On the other hand, a finite basis set could in fact give a faithful physical representation of the system if the system has a finite spectrum and the size of the basis is greater than or equal to the size of the spectrum. We will confirm this observation below in the case of potential (1) with $A_+ = 0$ where it becomes the Morse potential with a finite spectrum. We must, therefore, conclude that a finite basis set could be used to give an approximate representation of a physical system with an infinite spectrum like potential (1) and that the approximation improves by increasing the size of the basis. Unfortunately, we will find below that the size of the present basis set is constrained by physical requirements and could not be increased at will. On the other hand, it might be possible that a finite basis set could give an exact representation for part of the infinite spectrum whose size is less than or equal to the size of the basis set. Such a solution is called quasi-exact. By definition, a quasi-exact solution is a partial exact solution where part (not all) of the energy spectrum is obtained (see, for example, Ref. [21]). If the size of the basis is equal to the size of the quasi-exact solutions then the representation will be diagonal not tridiagonal. That is, the basis elements are in fact the eigenvectors of the associated Hamiltonian in the finite quasi-exact subspace.

In the atomic units, the time independent Schrödinger equation for the potential (1) is written in terms of the dimensionless variable $y$ as follows

$$\left[ y^2 \frac{d^2}{dy^2} + y \frac{d}{dy} - U(y) + \varepsilon \right] \psi(y) = 0, \tag{4}$$

where $U(y) = 2V(x)/\lambda^2 = \frac{1}{4} y^{-2} - A_- y^{-1} + A_+ y$ and $\varepsilon = 2E/\lambda^2$. If we write this wave equation as $\hat{J}\psi(y) = 0$ and substitute the ansatz (2) then finding the solution requires evaluating the action of the wave operator on the basis elements, $\hat{J}\phi_n(y)$. Now, the TRA dictates that this action must have the following tridiagonal structure [12]

$$\hat{J}\phi_n(y) = \omega(y)\left[ c_n \phi_n(y) + b_{n-1}\phi_{n-1}(y) + b_n \phi_{n+1}(y) \right], \tag{5}$$

where $\omega(y)$ is a non-zero analytic function on the whole real line and the coefficients $c_n$ and $b_n$ are $y$-independent and such that $b_n^2 > 0$ for all $n$. Substituting (2) into (4) and using (5), translates the problem into finding a solution to the following discrete algebraic equation

$$z P_n = a_n P_n + b_{n-1} P_{n-1} + b_n P_{n+1} = 0, \tag{6}$$

where we have written $c_n = a_n - z$ and $f_n = f_0 P_n$ making $P_0 = 1$. This is a three-tem recursion relation whose solution (due to $b_n^2 > 0$) is an orthogonal polynomial in $z$ provided that the recursion coefficients $a_n$ and $b_n$ are independent of $z$ [13,14]. It turns out that the polynomial argument $z$ depends on the energy and/or physical parameters of the problem. It was shown elsewhere that all the physical properties of the system (e.g., bound states energy spectrum, scattering phase shift, density of states, etc.) are obtained from the properties of this orthogonal polynomial $P_n(z)$ (its weight function, generating function, zeros, asymptotics, etc.) [12,15, 16]. It was also shown therein that the positive definite weight function of the polynomial $P_n(z)$ is $f_0^2(z)$. In the following section and by using the TRA tools, we obtain the recursion relation (6) associated with the potential function (1) and attempt to identify the corresponding orthogonal polynomial $P_n(z)$ wherefrom we obtain the physical properties of the system.

## 2. The TRA solution

Using the basis element (3) and after simple differential manipulation, we obtain

$$\hat{J}\phi_n(y) = G_n y^\alpha e^{-\beta/y} \left\{ y^2 \frac{d^2}{dy^2} + \left[ y(2\alpha+1) + 2\beta \right] \frac{d}{dy} \right.$$
$$\left. + \frac{1}{y}\beta(2\alpha-1) + \frac{\beta^2}{y^2} + \alpha^2 - U(y) + \varepsilon \right\} J_n^\mu(y) \tag{7}$$

Applying the differential equation of the Jacobi polynomial (A4) in Appendix A reduces this equation to the following form

$$\hat{J}\phi_n(y) = G_n y^\alpha e^{-\beta/y} \left\{ \left[ 2y\left(\alpha-\mu-\tfrac{1}{2}\right) + 2\beta - 1 \right] \frac{d}{dy} \right.$$
$$\left. + \frac{1}{y}\beta(2\alpha-1) + \frac{\beta^2}{y^2} + \alpha^2 - U(y) + n(n+2\mu+1) + \varepsilon \right\} J_n^\mu(y) \tag{8}$$

Choosing the basis parameters as $\alpha = \mu + \tfrac{1}{2}$ and $\beta = \tfrac{1}{2}$, will eliminate the derivative term and brings this equation into the following simple form

$$\hat{J}\phi_n(y) = G_n y^{\mu+\tfrac{1}{2}} e^{-1/2y} \left[ \frac{\mu}{y} + \frac{1/4}{y^2} + \left(n+\mu+\tfrac{1}{2}\right)^2 - U(y) + \varepsilon \right] J_n^\mu(y). \tag{9}$$

Now, compatibility of the TRA requirement (5) and the properties of the Bessel polynomial in Appendix A dictates that the expression inside the square brackets must be a linear function of $y$. Terms that are not linear must be eliminate by counter terms in $U(y)$. Therefore, the most general form of the function $U(y)$ that preserves the tridiagonal structure (5) is

$$U(y) = \frac{1/4}{y^2} + \frac{U_0}{y} + U_1 y, \tag{10}$$

such that the basis parameter $\mu$ is chosen as $\mu = U_0$ whereas $U_1$ is an arbitrary real parameter. In fact, this is the exact from of our proposed potential (1) with $U_0 = -A_-$ and $U_1 = A_+$. Note that any constant in the potential function $U(y)$ could be absorbed into the energy $\varepsilon$ by a simple redefinition. With this expression of $U(y)$, Eq. (9) becomes

$$\hat{J}\phi_n(x) = G_n y^{\mu+\tfrac{1}{2}} e^{-1/2y} \left[ -A_+ y + \left(n+\mu+\tfrac{1}{2}\right)^2 + \varepsilon \right] J_n^\mu(y), \tag{11}$$

where $\mu = -A_-$. The first term $-A_+ y$ is evaluated using the recursion relation of the Bessel polynomial (A2) in Appendix A giving

$$\hat{J}\phi_n(x) = \omega(x) \left\{ \left[ \frac{-2\mu}{(n+\mu)(n+\mu+1)} - \frac{4}{A_+}\left(n+\mu+\tfrac{1}{2}\right)^2 - \frac{4\varepsilon}{A_+} \right] \phi_n(x) \right.$$
$$\left. - \frac{n}{(n+\mu)\left(n+\mu+\tfrac{1}{2}\right)} \frac{G_n}{G_{n-1}} \phi_{n-1}(x) + \frac{n+2\mu+1}{(n+\mu+1)\left(n+\mu+\tfrac{1}{2}\right)} \frac{G_n}{G_{n+1}} \phi_{n+1}(x) \right\} \tag{12}$$

where $\omega(x) = -\tfrac{1}{4}A_+$. This is identical to (5) if we write

$$c_n = \frac{-2\mu}{(n+\mu)(n+\mu+1)} - \frac{4}{A_+}\left(n+\mu+\tfrac{1}{2}\right)^2 - \frac{4\varepsilon}{A_+}, \tag{13a}$$

$$b_n = \frac{-1}{n+\mu+1}\sqrt{\frac{-(n+1)(n+2\mu+1)}{\left(n+\mu+\tfrac{1}{2}\right)\left(n+\mu+\tfrac{3}{2}\right)}}. \tag{13b}$$

Note that due to the restricted range of $\mu$ as $\mu < -N - \tfrac{1}{2}$, the term under the square root in (13b) is always positive making $b_n^2 > 0$ for $n = 0,1,2,\ldots,N-1$. If we write $c_n = a_n - z = a_n - (4\varepsilon/A_+)$ then we obtain the three-term recursion relation (6) for the polynomial $P_n(4\varepsilon/A_+)$. To identify this orthogonal polynomial, which contains all the physical properties of the system, we start by recasting its recursion relation into a standard format through defining the polynomial $B_n^\mu(z;\gamma)$ via $P_n(z) = (G_n/G_0) B_n^\mu(z;\gamma)$ where $\gamma = -4/A_+$ and $G_n$ is given below Eq. (3). This definition results in the following recursion relation for $B_n^\mu(z;\gamma)$

$$z B_n^\mu(z;\gamma) = \left[\frac{-2\mu}{(n+\mu)(n+\mu+1)} + \gamma\left(n+\mu+\tfrac{1}{2}\right)^2\right] B_n^\mu(z;\gamma)$$
$$- \frac{n}{(n+\mu)\left(n+\mu+\tfrac{1}{2}\right)} B_{n-1}^\mu(z;\gamma) + \frac{n+2\mu+1}{(n+\mu+1)\left(n+\mu+\tfrac{1}{2}\right)} B_{n+1}^\mu(z;\gamma) \tag{14}$$

Now, this recursion relation gives all of the polynomials $\{B_n^\mu(z;\gamma)\}_{n=o}^{N-1}$ explicitly starting with $B_0^\mu(z;\gamma) = 1$ and setting $B_{-1}^\mu(z;\gamma) = 0$. Note that (14) differs significantly from (A2) for the Bessel polynomial due to the presence of $\left(n+\mu+\tfrac{1}{2}\right)^2$ in the diagonal term of the recursion. However, if $\gamma = 0$ then $B_n^\mu(z;0) = J_n^\mu(z/4)$. All of our attempts to compare the polynomial $B_n^\mu(z;\gamma)$ with known ones failed to produce a match. We tried using the table of recurrence formulas in Chihara's book [13] and the properties of the hypergeometric orthogonal polynomials in the book by Koekoek *et. al* [17]. We also looked at the Chapter on Orthogonal Polynomials in the Digital Library of Mathematical Functions [18] and compared with the information available in CAOP (Computer Algebra & Orthogonal Polynomials) [19]. We also tried using computer algebra systems (such as **rec2ortho** of Koornwinder and Swarttouw, or **retode** of Koepf and Schmersau [20]) that identify the polynomials from their recurrence relations. Consequently, we were forced to resort to numerical means to extract the physical information from $B_n^\mu(z;\gamma)$ as will be shown in the following section.

Finally, our partial solution of the Schrödinger equation with the potential (1) reads as follows

$$\psi(x) = \sqrt{\rho(4\varepsilon/A_+)} \sum_{n=0}^{N-1} B_n^\mu(4\varepsilon/A_+; -4/A_+) \phi_n(x), \tag{15}$$

where the basis element $\phi_n(x)$ is given by Eq. (3) with $\mu = -A_-$, $\alpha = \mu + \tfrac{1}{2}$ and $\beta = \tfrac{1}{2}$. Moreover, $\rho(z)$ is the positive definite weight function for the polynomial $B_n^\mu(z;\gamma)$. Note that the condition $\mu < -N - \tfrac{1}{2}$ implies that the number of bound states $N$ obtained by the TRA is the largest integer less than or equal to $A_- - \frac{1}{2}$. However, the actual number of bound states associated with this confining potential is clearly infinite. Thus, our TRA solution obtained here is (at best) a quasi-exact solution of the problem. In the following section, we calculate the lowest bound state energies and plot the corresponding wavefunctions.

## 2. Results and discussion

The first straight-forward and almost-trivial result holds when $A_+ = 0$. As seen from Eq. (11), the representation in this case becomes diagonal not tridiagonal and the energy spectrum becomes $\varepsilon_n = -\left(n + \mu + \frac{1}{2}\right)^2$. That is,

$$E_n = -\frac{1}{2}\lambda^2 \left(n - A_- + \frac{1}{2}\right)^2, \tag{16}$$

where $n = 0, 1, 2, .., A_- - \frac{1}{2}$ which means that bound states exist only if $A_- \geq \frac{1}{2}$. This is the well-known result for the Morse potential in one dimension [9,22]. The corresponding bound state wavefunction is

$$\psi_n(x) = \phi_n(x) = G_n y^{\frac{1-A_-}{2}} e^{-1/2y} J_n^{-A_-}(y), \tag{17}$$

Using the orthogonality of the Bessel polynomial (A3) in Appendix A, we can easily show that $\psi_n(x)$ is orthonormal. That is,

$$\langle \psi_n | \psi_m \rangle = \lambda \int_{-\infty}^{+\infty} \phi_n(x)\phi_m(x)dx = G_n G_m \int_0^\infty y^{-2A_-} e^{-1/y} J_n^{-A_-}(y) J_m^{-A_-}(y) dy = \delta_{nm}. \tag{18}$$

Therefore, the finite incomplete basis set whose elements are given by Eq. (3) does give a faithful physical representation for the finite number of bound states of the system whose potential function is given by Eq. (1) with $A_+ = 0$. Now for the general case where $A_+ \neq 0$, the TRA asset is in the polynomial $B_n^\mu(4\varepsilon/A_+; -1/A_+)$ that contains all physical properties of the system [12]. Unfortunately, as noted above, the analytic properties of this polynomial are not found in the mathematics literature. It remains an open problem in orthogonal polynomials along with other similar ones. For a discussion on these open problems, one may consult References [23-25]. Note that this shortcoming is in spite of the fact that the recursion relation (14) gives all of these polynomials explicitly, albeit not in closed form, to any desired degree starting with $B_0^\mu(z; \gamma) = 1$. Nonetheless, we will be able to obtain a very stable and convergent numerical result. For example, the Hamiltonian representation in the "Bessel basis" (3) is a finite $N \times N$ tridiagonal symmetric matrix, which is obtained from the three-term recursion relation of $P_n(4\varepsilon/A_+)$ as

$$H_{n,m} = \frac{\lambda^2}{8} A_+ \left(a_n \delta_{n,m} + b_{n-1} \delta_{n,m+1} + b_n \delta_{n,m-1}\right), \tag{19}$$

where $a_n$ and $b_n$ are obtained from (13a) and (13b) with $n,m = 0,1,..,N$. Diagonalizing this matrix will give the lowest bound states energy spectrum. Table 1 is a list of these bound states energies for several values of $A_-$ and fixed $A_+ = 2$, whereas in Table 2 we fix $A_- = 6$ and vary $A_+$. The first column in Table 2 represents an accuracy check where we reproduced the well-known energy spectrum of the Morse potential (16). However, we like to reiterate that the solution obtained here is at best a quasi-exact solution; not an exact solution. That is, we obtained only a finite part of the energy spectrum not the whole infinite spectrum (except, of course, when $A_+ = 0$ in which case the whole spectrum is finite). Moreover, due to the limited size of the Hamiltonian matrix (19), because $(n,m) \leq N-1$, we expect that the accuracy of our results be substantially reduced for higher energy levels. In Appendix B, we give an independent evaluation of the energy spectrum by diagonalizing the Hamiltonian matrix in a complete square integrable basis. Numerically, however, we choose a relatively large subset of this basis (i.e., large matrix size) to produce numbers that are more accurate. Using the physical parameters in Table 1 and Table 2, we reproduce these as Table 3 and Table 4, respectively. Comparison of these results confirms our expectation that the accuracy in Table 1 and Table 2 is substantially reduced for higher energy levels. These findings clearly stress the importance and urgency for deriving the analytic properties of the orthogonal polynomial $B_n^\mu(z;\gamma)$ so that we could obtain an accurate quasi-exact solution of the problem.

Finally, we can use (15) to calculate the bound states wavefunctions. Figure 2 is such a plot showing the un-normalized wavefunctions corresponding to the lowest part of the spectrum $\{\varepsilon_m\}$ in the first column of Table 3, which are calculated using

$$\psi_m(x) = e^{\left(\frac{1}{2}-A_-\right)\lambda x} \exp\left(-\tfrac{1}{2}e^{-\lambda x}\right) \sum_n G_n B_n^{-A_-}(4\varepsilon_m/A_+;-4/A_+) J_n^{-A_-}(e^{\lambda x}). \tag{20}$$

To understand the features of the wavefunction, we superimposed the potential (not to scale) on the same plots and indicated the corresponding energy level by a horizontal line. The wave function plots show interesting confinement features of the potential. Firstly, the wavefunction does not diminish abruptly at the walls of the potential well but exhibits some penetration with a short decay tail; shorter on the left than on the right due to the stronger confinement of $e^{-2\lambda x}$ on the left relative to $e^{\lambda x}$ on the right. Secondly, the apparent bunching of the wavefunction for low energies on the left side of the well where it is deeper. Thirdly, at higher energies, the particle does not "feel" the topography at bottom of the well exhibiting oscillations. However, at these higher energies, the particle still feels the weaker right wall and tends to cluster there.

## Acknowledgements

We are grateful to A. J. Sous (Al-Quds Open University) and I. A. Assi (Memorial University of Newfoundland) for pointing out the diminished accuracy of higher energy levels in Tables 1 and 2.

# Appendix A: The Bessel polynomial on the real line

The Bessel polynomial on the real line is defined in terms of the hypergeometric or confluent hypergeometric functions as follows (see section 9.13 of the book by Koekoek *et. al* [17]):

$$J_n^\mu(x) = {}_2F_0\!\left(\begin{matrix}-n, n+2\mu+1\\ -\end{matrix}\Big| -x\right) = (n+2\mu+1)_n\, x^n\, {}_1F_1\!\left(\begin{matrix}-n\\ -2(n+\mu)\end{matrix}\Big| 1/x\right), \tag{A1}$$

where $x \geq 0$, $n = 0,1,2,..,N$ and $N$ is a non-negative integer. The real parameter $\mu$ is negative such that $\mu < -N - \frac{1}{2}$. The Pochhammer symbol $(a)_n$ (a.k.a. shifted factorial) is defined as $a(a+1)(a+2)...(a+n-1) = \frac{\Gamma(n+a)}{\Gamma(a)}$. The Bessel polynomial could also be written in terms of the associated Laguerre polynomial with discrete index as: $J_n^\mu(x) = n!(-x)^n L_n^{-(2n+2\mu+1)}(1/x)$. The three-term recursion relation reads as follows:

$$2x J_n^\mu(x) = \frac{-\mu}{(n+\mu)(n+\mu+1)} J_n^\mu(x)$$
$$-\frac{n}{(n+\mu)(2n+2\mu+1)} J_{n-1}^\mu(x) + \frac{n+2\mu+1}{(n+\mu+1)(2n+2\mu+1)} J_{n+1}^\mu(x) \tag{A2}$$

Note that the constraints on $\mu$ and on the maximum polynomial degree make this recursion definite (i.e., the signs of the two recursion coefficients multiplying $J_{n\pm1}^\mu(x)$ are the same). Otherwise, these polynomials could not be orthogonal on the real line but on the unit circle in the complex plane. The orthogonality relation reads as follows

$$\int_0^\infty x^{2\mu} e^{-1/x} J_n^\mu(x) J_m^\mu(x)\, dx = -\frac{n!\,\Gamma(-n-2\mu)}{2n+2\mu+1}\, \delta_{nm}. \tag{A3}$$

The differential equation is

$$\left\{ x^2 \frac{d^2}{dx^2} + \left[1 + 2x(\mu+1)\right] \frac{d}{dx} - n(n+2\mu+1) \right\} J_n^\mu(x) = 0. \tag{A4}$$

The forward shift differential relation reads as follows

$$\frac{d}{dx} J_n^\mu(x) = n(n+2\mu+1) J_{n-1}^{\mu+1}(x). \tag{A5}$$

On the other hand, the backward shift differential relation is

$$x^2 \frac{d}{dx} J_n^\mu(x) = -(2\mu x + 1) J_n^\mu(x) + J_{n+1}^{\mu-1}(x). \tag{A6}$$

We can write $J_{n+1}^{\mu-1}(x)$ in terms of $J_n^\mu(x)$ and $J_{n\pm1}^\mu(x)$ as follows

$$2J_{n+1}^{\mu-1}(x) = \frac{(n+1)(n+2\mu)}{(n+\mu)(n+\mu+1)} J_n^{\mu}(x)$$
$$+ \frac{n(n+1)}{(n+\mu)(2n+2\mu+1)} J_{n-1}^{\mu}(x) + \frac{(n+2\mu)(n+2\mu+1)}{(n+\mu+1)(2n+2\mu+1)} J_{n+1}^{\mu}(x) \tag{A7}$$

Using this identity and the recursion relation (A2), we can rewrite the backward shift differential relation as follows

$$2x^2 \frac{d}{dx} J_n^{\mu}(x) = n(n+2\mu+1) \times$$
$$\left[ -\frac{J_n^{\mu}(x)}{(n+\mu)(n+\mu+1)} + \frac{J_{n-1}^{\mu}(x)}{(n+\mu)(2n+2\mu+1)} + \frac{J_{n+1}^{\mu}(x)}{(n+\mu+1)(2n+2\mu+1)} \right] \tag{A8}$$

The generating function is

$$\sum_{n=0}^{\infty} J_n^{\mu}(x) \frac{t^n}{n!} = \frac{2^{2\mu}}{\sqrt{1-4xt}} \left(1+\sqrt{-4xt}\right)^{-2\mu} \exp\left[2t/(1+\sqrt{1-4xt})\right]. \tag{A9}$$

## Appendix B: Alternative evaluation of the energy spectrum

In this appendix, we obtain an independent and more accurate numerical evaluation of the energy spectrum associated with the exponential potential well (1) by diagonalizing the Hamiltonian matrix in an appropriate discrete square integrable basis. We choose the following elements for the basis set, which is called the "Laguerre basis",

$$\phi_n(x) = A_n y^{\alpha} e^{-y/2} L_n^{\nu}(y), \tag{B1}$$

where $L_n^{\nu}(y)$ is the Laguerre polynomial and the normalization constant is conveniently chosen as $A_n = \sqrt{n!/\Gamma(n+\nu+1)}$. However, the coordinate transformation is chosen as $y(x) = e^{-\lambda x}$, which is the inverse of that in the "Bessel basis" (3). The real parameters $\alpha$ and $\nu$ will be chosen below but with $\nu > -1$. In the atomic units $\hbar = M = 1$, the Hamiltonian operator is

$$H = -\frac{1}{2} \frac{d^2}{dx^2} + V(x) = -\frac{\lambda^2}{2} \left[ y^2 \frac{d^2}{dy^2} + y \frac{d}{dy} - W(y) \right], \tag{B2}$$

where $W(y) = 2V(x)/\lambda^2 = \frac{1}{4} y^2 - A_- y + A_+ y^{-1}$. Using the differential equation of the Laguerre polynomial, $\left[ y \frac{d^2}{dy^2} + (\nu+1-y) \frac{d}{dy} + n \right] L_n^{\nu}(y) = 0$, we obtain the following action of $H$ on the basis elements

$$-\frac{2}{\lambda^2} H |\phi_n\rangle = A_n y^{\alpha} e^{-y/2} \left[ (2\alpha-\nu) y \frac{d}{dy} + \frac{y^2}{4} - \left(n+\alpha+\tfrac{1}{2}\right) y + \alpha^2 - W(y) \right] |L_n^{\nu}\rangle. \tag{B3}$$

Using the differential property of the Laguerre polynomial, $y \frac{d}{dy} L_n^{\nu} = n L_n^{\nu} - (n+\nu) L_{n-1}^{\nu}$, and the explicit expression for $W(y)$ in this equation, we obtain

–9–

$$-\frac{2}{\lambda^2} H |\phi_n\rangle = A_n y^\alpha e^{-y/2} \times$$
$$\left\{ (\nu - 2\alpha)(n+\nu) |L_{n-1}^\nu\rangle + \left[ \alpha^2 + n(2\alpha - \nu) - \left(n + \alpha + \tfrac{1}{2} - A_-\right) y - \frac{A_+}{y} \right] |L_n^\nu\rangle \right\} \quad \text{(B4)}$$

Using the integral measure $\lambda \int_{-\infty}^{+\infty} dx = \int_0^\infty y^{-1} dy$, we obtain the following matrix elements of the Hamiltonian in the Laguerre basis (B1)

$$-\frac{2}{\lambda^2} \langle \phi_m | H | \phi_n \rangle = A_m A_n (\nu - 2\alpha)(n+\nu) \langle L_m^\nu | y^{2\alpha-1} e^{-y} | L_{n-1}^\nu \rangle$$
$$+ A_m A_n \langle L_m^\nu | y^{2\alpha-1} e^{-y} \left[ \alpha^2 + n(2\alpha-\nu) - \left(n+\alpha+\tfrac{1}{2}-A_-\right)y - \frac{A_+}{y} \right] | L_n^\nu \rangle \quad \text{(B5)}$$

If we define $\langle m | f(y) | n \rangle = A_m A_n \langle L_m^\nu | y^\nu e^{-y} f(y) | L_n^\nu \rangle = A_m A_n \int_0^\infty y^\nu e^{-y} f(y) L_n^\nu(y) L_m^\nu(y) dy$ then we obtain

$$-\frac{2}{\lambda^2} \langle \phi_m | H | \phi_n \rangle = (\nu - 2\alpha) \sqrt{n(n+\nu)} \langle m | y^{2\alpha-\nu-1} | n-1 \rangle$$
$$+ \langle m | y^{2\alpha-\nu-1} \left[ \alpha^2 + n(2\alpha-\nu) - \left(n+\alpha+\tfrac{1}{2}-A_-\right)y - \frac{A_+}{y} \right] | n \rangle \quad \text{(B6)}$$

Using the orthogonality of the Laguerre polynomial, $A_m A_n \int_0^\infty y^\nu e^{-y} L_m^\nu(y) L_n^\nu(y) dx = \delta_{m,n}$, and choosing $2\alpha = \nu + 1$, we obtain

$$-\frac{2}{\lambda^2} \langle \phi_m | H | \phi_n \rangle = -\sqrt{n(n+\nu)} \delta_{m,n-1} + \left[ n + \tfrac{1}{4}(\nu+1)^2 \right] \delta_{m,n}$$
$$- \left(n + \tfrac{\nu}{2} + 1 - A_-\right) \langle m | y | n \rangle - A_+ \langle m | y^{-1} | n \rangle \quad \text{(B7)}$$

Using the three-term recursion relation of the Laguerre polynomials $y L_n^\nu = (2n+\nu+1) L_n^\nu - (n+\nu) L_{n-1}^\nu - (n+1) L_{n+1}^\nu$ and their orthogonality, we obtain the following tridiagonal symmetric matrix representation for $\langle m | y | n \rangle$

$$\langle m | y | n \rangle = (2n+\nu+1) \delta_{m,n} - \sqrt{n(n+\nu)} \delta_{m,n-1} - \sqrt{(n+1)(n+\nu+1)} \delta_{m,n+1} := T_{m,n}. \quad \text{(B8)}$$

On the other hand, we can use any integral approximation routine to evaluate the integral $\langle m | y^{-1} | n \rangle$. If we use the Gauss quadrature [26-29] associated with the Laguerre polynomial, then we obtain

$$\langle m | y^{-1} | n \rangle \cong \sum_{k=0}^{K-1} (1/\xi_k) \Lambda_{m,k} \Lambda_{n,k}, \quad \text{(B9)}$$

where $\{\xi_k\}_{k=0}^{K-1}$ is the set of eigenvalues of the $K \times K$ tridiagonal symmetric matrix $T$ defined in (B8) and $\{\Lambda_{m,k}\}_{m=0}^{K-1}$ is its normalized eigenvector associated with the eigenvalue $\xi_k$.

Now, for any given set of potential parameters $\{\lambda, A_\pm\}$, we only need to assign a value to the basis parameter $\nu$ to obtain the $K \times K$ matrix representation of the Hamiltonian $H$ in (B7). We should expect that the physical results are independent of any choice of value for $\nu$. To proceed, we choose a value for $\nu$ from within a range, called the "plateau of stability", where the results obtained have no significant deviations (within the desired accuracy) from the exactly known energy spectrum of the Morse potential for $A_+ = 0$. The plateau widens with the size of the matrix $H$. That is, the range from which one can choose a "good" value for $\nu$ increases with $K$. Ideally, the size of the plateau becomes infinite if $K \to \infty$. That is, the results do not change with $\nu$ as long as $\nu > -1$. As a result, we choose the value $\nu = 0$ from within the plateau. Note also that with $2\alpha = \nu + 1$ the basis becomes an orthonormal set (i.e., $\langle \phi_m | \phi_n \rangle = \delta_{m,n}$). We could have also obtained the energy spectrum for different relations between the basis parameters $\alpha$ and $\nu$. For example, we could have also chosen $2\alpha = \nu$ or $2\alpha = \nu + 2$ with the corresponding basis overlap matrix $\langle \phi_m | \phi_n \rangle = \langle m | y^{-1} | n \rangle$ and $\langle \phi_m | \phi_n \rangle = \langle m | y | n \rangle$, respectively. We opted for the choice $2\alpha = \nu + 1$ because it produced the most accurate match with the exact energy spectrum of the Morse potential relative to the other two choices for the same basis size.

Finally, using the physical parameters in Table 1 and Table 2, we reproduce them as Table 3 and Table 4, respectively, using the procedure described in this Appendix (with $2\alpha = \nu + 1$ and $K = 100$).

## References:

[1] N. Zettili, *Quantum Mechanics: Concepts and Applications*, 2nd ed. (Wiley, New York, 2009)

[2] P. L. Ferreira, J. A. Helayel, and N. Zagury, *A Linear-Potential Model for Quark Confinement*, Il Nuovo Cimento **55** (1980) 215

[3] A. Nakamura and T. Saito, *QCD color interactions between two quarks*, Phys. Lett. B **621** (2005) 171

[4] F. Cooper, A. Khare and U. Sukhatme, *Supersymmetry in Quantum Mechanics* (World Scientific, Singapore, 2004).

[5] M. Bander and C. Itzykson, *Group Theory and the Hydrogen Atom*, Rev. Mod. Phys. **38** (1966) 330

[6] Y. Alhassid, F. Iachello and F. Gursey, *Group theory of the Morse oscillator*, Chem. Phys. Lett. **99** (1983) 27

[7] L. Infeld and T. D. Hull, *The Factorization Method*, Rev. Mod. Phys. **23** (1951) 21

[8] H. Ciftci, R. L. Hall, and N. Saad, *Construction of exact solutions to eigenvalue problems by the asymptotic iteration method*, J. Phys. A **38** (2005) 1147

[9] De R., Dutt R., Sukhatme U., *Mapping of shape invariant potentials under point canonical transformations*, J. Phys. A **25** (1992), L843.

[10] R. P. Feynman and A.R. Hibbs, *Quantum Mechanics and Path Integrals* (McGraw Hill, New York, 1965).

[11] A. F. Nikiforov and V. B. Uvarov, *Special Functions of Mathematical Physics* (Basel, Birkhauser, 1988).

[12] A. D. Alhaidari and H. Bahlouli, *Tridiagonal Representation Approach in Quantum Mechanics*, Phys. Scripta **94** (2019) 125206

[13] T. S. Chihara, *An introduction to orthogonal polynomials,* Mathematics and its Applications, Vol. 13, (Gordon and Breach, New York - London - Paris, 1978)

[14] G. Szegő, *Orthogonal Polynomials*, 4th ed. (American Mathematical Society, Providence, 1975)

[15] A. D. Alhaidari and M. E. H. Ismail, *Quantum mechanics without potential function*, J. Math. Phys. **56** (2015) 072107

[16] A. D. Alhaidari, *Representation of the quantum mechanical wavefunction by orthogonal polynomials in the energy and physical parameters*, Commun. Theor. Phys. **72** (2020) 015104

[17] R. Koekoek, P. A. Lesky and R. F. Swarttouw: *Hypergeometric Orthogonal Polynomials and Their q-Analogues* (Springer, Heidelberg 2010)

[18] F. W. J. Olver, A. B. Olde Daalhuis, D. W. Lozier, B. I. Schneider, R. F. Boisvert, C. W. Clark, B. R. Miller, and B. V. Saunders (Editors), *NIST Digital Library of Mathematical Functions*, Release 1.0.22 of 2019-03-15 (http://dlmf.nist.gov).

[19] CAOP: Computer Algebra & Orthogonal Polynomials, http://www.caop.org/ (maintained by Wolfram Koepf, University of Kassel)

[20] W. Koepf and D. Schmersau, *Recurrence equations and their classical orthogonal polynomial solutions*, Appl. Math. Comput. **128** (2002) 303

[21] A. G. Ushveridze, *Quasi-Exactly Solvable Models in Quantum Mechanics* (Institute of Physics Publishing, Bristol, 1994)

[22] P. C. Ojha, *SO*(2,1) *Lie algebra, the Jacobi matrix and the scattering states of the Morse oscillator*, J. Phys. A **21** (1988) 875

[23] A. D. Alhaidari, *Open problem in orthogonal polynomials*, Rep. Math. Phys. **84** (2019) 393

[24] A. D. Alhaidari, *Orthogonal polynomials derived from the tridiagonal representation approach*, J. Math. Phys. **59** (2018) 013503

[25] W. Van Assche, *Solution of an open problem about two families of orthogonal polynomials*, SIGMA **15** (2019) 005

[26] Appendix A in: A. D. Alhaidari, *Reconstructing the potential function in a formulation of quantum mechanics based on orthogonal polynomials*, Commun. Theor. Phys. **68** (2017) 711

[27] G. H. Golub and G. Meurant, *Matrices, Moments and Quadrature with Applications* (Princeton University Press, 2010)

[28] W. Gautschi, *Orthogonal polynomials: computation and approximation* (Oxford University Press, 2004)

[29] P. J. Davis and P. Rabinowitz, *Methods of numerical integration*, 2nd ed. (Academic Press, 1984)

## Tables Captions:

**Table 1**: The lowest bound states energies (in atomic units) corresponding to $A_+ = 2$ and for several values of $A_-$ as indicated. We took $\lambda = 1$.

**Table 2**: The lowest bound states energies (in atomic units) corresponding to $A_- = 6$ and for several values of $A_+$ as indicated. We took $\lambda = 1$. The first column is an accuracy check where we reproduced the well-known energy spectrum of the Morse potential.

**Table 3**: A reproduction of Table 1 using the procedure outlined in Appendix B. We took the Hamiltonian matrix size 100×100.

**Table 4**: A reproduction of Table 2 using the procedure outlined in Appendix B. We took the Hamiltonian matrix size 100×100.

## Figures Captions:

**Fig. 1**: The potential function (1) for $\lambda = 1$, $A_+ = 2$, and several values of $A_-$. The six traces from top to bottom correspond to $A_- = -4, 0, 4, 6, 8, 10$.

**Fig. 2**: The un-normalized wavefunctions $\{\psi_n(x)\}_{n=0}^{n=5}$ corresponding to the energies shown in the first column of Table 3 where $A_+ = 2$ and $A_- = 8$. The potential function (dashed curve) is superimposed on the same plot (not to scale) and the corresponding energy level is indicated by the horizontal dotted line.

**Table 1**

| $n$ | $A_- = 8$ | $A_- = 6$ | $A_- = 4$ |
|---|---|---|---|
| 0 | -28.053627 | -15.025220 | -5.960092 |
| 1 | -21.029931 | -9.975990 | -2.808535 |
| 2 | -14.992219 | -5.880414 | -0.106373 |
| 3 | -9.927105 | -2.662228 | |
| 4 | -5.801982 | 0.418853 | |
| 5 | -2.518657 | | |
| 6 | 0.948521 | | |

**Table 2**

| $n$ | $A_+ = 0$ | $A_+ = 4$ | $A_+ = 8$ | $A_+ = 12$ |
|---|---|---|---|---|
| 0 | -15.125000 | -14.925872 | -14.728422 | -14.532562 |
| 1 | -10.125000 | -9.828860 | -9.539719 | -9.256701 |
| 2 | -6.125000 | -5.645103 | -5.195514 | -4.765950 |
| 3 | -3.125000 | -2.229669 | -1.374691 | -0.509143 |
| 4 | -1.125000 | 2.004504 | 5.213347 | 8.439356 |

**Table 3**

| $n$ | $A_- = 8$ | $A_- = 6$ | $A_- = 4$ |
|---|---|---|---|
| 0 | -28.053627 | -15.025220 | -5.960092 |
| 1 | -21.029931 | -9.975990 | -2.809728 |
| 2 | -14.992219 | -5.880416 | -0.410359 |
| 3 | -9.927105 | -2.667212 | 1.581097 |
| 4 | -5.801990 | -0.146643 | 3.496007 |
| 5 | -2.530695 | 2.014391 | 5.470629 |
| 6 | 0.093216 | 4.103609 | 7.493054 |
| 7 | 2.391258 | 6.224812 | 9.682386 |

**Table 4**

| $n$ | $A_+ = 0$ | $A_+ = 4$ | $A_+ = 8$ | $A_+ = 12$ |
|---|---|---|---|---|
| 0 | -15.125000 | -14.925872 | -14.728422 | -14.532562 |
| 1 | -10.125000 | -9.828860 | -9.539720 | -9.256705 |
| 2 | -6.125000 | -5.645149 | -5.196474 | -4.770689 |
| 3 | -3.125000 | -2.262598 | -1.542214 | -0.896463 |
| 4 | -1.125000 | 0.549743 | 1.669961 | 2.607619 |
| 5 | -0.125000 | 3.095928 | 4.694312 | 5.962793 |

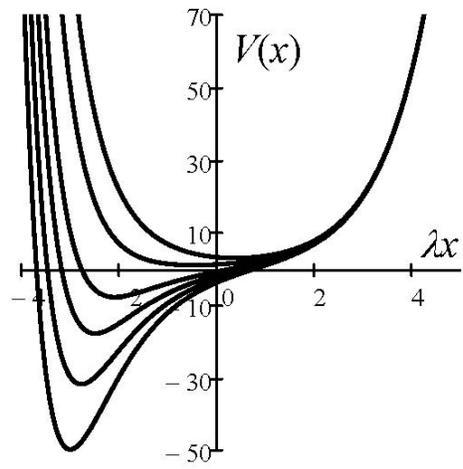

**Fig. 1**

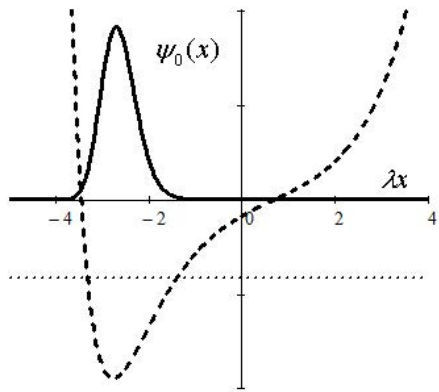 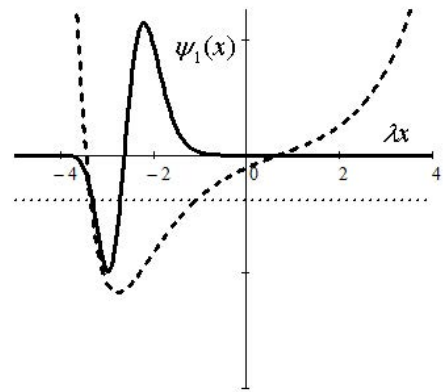

**Fig. 2a**  **Fig. 2b**

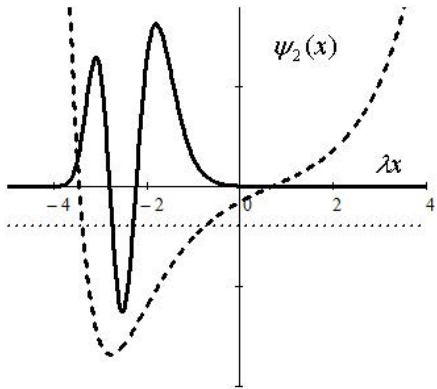 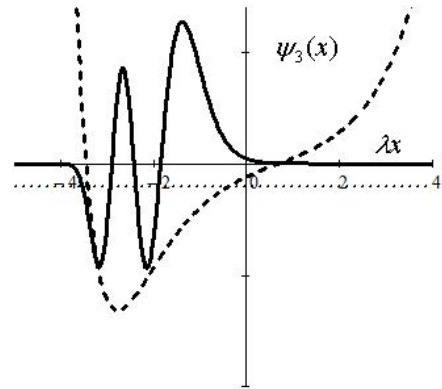

**Fig. 2c**  **Fig. 2d**

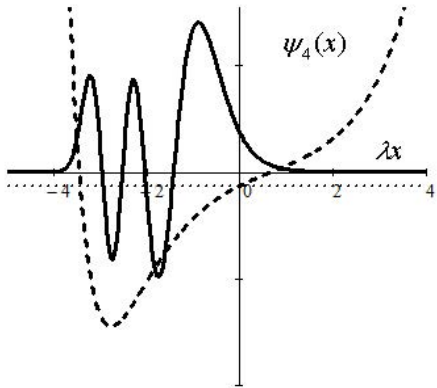 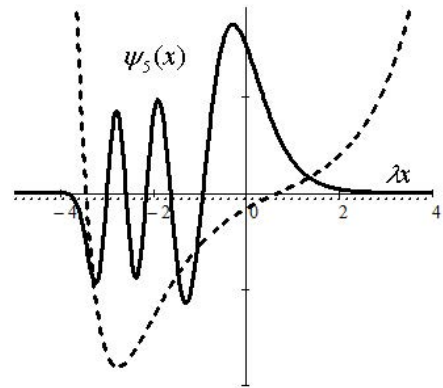

**Fig. 2e**  **Fig. 2f**